\documentstyle[aps, pra, twocolumn, epsfig]{revtex}

\def\bea{\begin{eqnarray}}
\def\eea{\end{eqnarray}}
\def\be{\begin{equation}}
\def\ee{\end{equation}}
\begin{document}


\author{Bogdan Damski and Krzysztof Sacha}
\address{
Instytut Fizyki imienia Mariana Smoluchowskiego,
  Uniwersytet Jagiello\'nski,\\
 ulica Reymonta 4, PL-30-059 Krak\'ow, Poland \\
}
\title{
Changes of the topological charge of vortices
}

\maketitle

\begin{abstract}
We consider changes of the topological charge 
of vortices in quantum mechanics by investigating 
analytical examples where the creation or annihilation of vortices occurs.
In classical hydrodynamics of non-viscous fluids
the Helmholtz-Kelvin theorem ensures that the velocity field circulation is
conserved. We discuss applicability of the theorem in the hydrodynamical
formulation of  quantum mechanics 
showing that the assumptions of the theorem may be broken in quantum evolution of
the wavefunction leading to a change of
the topological charge.
\end{abstract}


\section{Introduction}
Vortices can be found both in classical and quantum physics.
One can encounter vortices, for instance, 
in water  that spins around or in the air 
as a ring of smoke \cite{feyn}. 
Quantum mechanics can be formulated in a language of 
hydrodynamics
(see e.g. \cite{ghosh,bial92} and references therein). 
Such a formulation 
(very useful also in quantum chemistry \cite{ghosh}) 
provides a basis for definition of topological defects like vortices
~\cite{bial00}, whose features are even more striking than 
those that we find in classical physics.
Quantum vortices appear not only in systems
described by the linear Schr\"odinger equation. 
Indeed, they were experimentally observed 
in superfluid HeII \cite{andron} and a Bose-Einstein condensate (BEC) 
of trapped alkali atoms ~\cite{cornell,dalibard} which are typically
described, within the mean-field approximation,
by a nonlinear equation \cite{pitaevskii}.

In hydrodynamics of ordinary non-viscous fluids, the
circulation of the velocity field is conserved in time evolution due to 
the celebrated Helmholtz-Kelvin theorem (HKT) \cite{landau}. 
This theorem is often employed in quantum mechanics
\cite{bial92,bial00,bolda,anderson}.
However, uncritical usage of the HKT, 
may lead to the incorrect conclusion that 
stability of vortices and vortex rings in quantum fluids
is fully guaranteed by the HKT \cite{anderson}. 
In the present paper we discuss the basic assumptions of the HKT 
and show analytical examples where these assumptions can be easily broken 
by quantum evolution of the wave function, leading to changes of the topological 
charge. Difficulties in fulfilling the assumptions of the HKT have been already
pointed out in conclusions of Ref.~\cite{koplik}. 

\section{Hydrodynamical formulation of quantum mechanics}
To establish connections between quantum mechanics and fluid 
dynamics we write the wave function in the form
$\Psi(\vec{r}, t)=\sqrt{\rho(\vec{r}, t)} \exp(i \chi(\vec{r}, t))$ 
\cite{ghosh},
(where $\rho(\vec{r}, t)$ stands for density of a probability fluid)
and define the velocity field 
\begin{equation}
\label{velocity}
\vec{v} = \frac{\hbar}{m} \vec{\nabla}\chi(\vec{r}, t).
\end{equation}
Provided $\vec v$ and its partial derivatives are well defined, 
we may rewrite
the Schr\"odinger equation [with a potential $V(\vec{r}, t)$] 
in the form
\begin{equation}
\label{dyna}
 m \frac{\partial \vec{v}}{\partial t} + \vec{\nabla}\left(\frac{1}{2} m v^2 + 
V - \frac{\hbar^2}{2 m} 
\frac{\nabla^2 \sqrt{\rho}}{\sqrt{\rho}}\right) = 0,
\end{equation}
\begin{equation}
\label{cont}
\frac{\partial \rho}{\partial t} + \vec{\nabla}\cdot\left(\vec{v}\rho\right) =0.
\end{equation}
Equation (\ref{cont}) is an ordinary continuity equation, while 
Eq.~(\ref{dyna}), in the limit of $\hbar \rightarrow 0$, becomes
similar to
the dynamical equation of a curl-free non-viscous fluid
\cite{landau}.

In quantum mechanics 
the wave function has to be single valued. To satisfy this condition 
one arrives at
the Feynman-Onsager quantization condition \cite{feyn1} for the circulation 
of the velocity field $\Gamma_C$ around any closed contour $C$
\begin{equation}
\label{circ}
\Gamma_C = \oint_{C} \vec{v} \cdot d\vec{l} = n \frac{2\pi\hbar}{m},
\end{equation}
where  $n = 0, \pm 1, \pm 2, \dots$. 
Due to the definition (\ref{velocity}), we may not expect $n\ne0$
unless at a certain point on a surface encircled by the contour $C$, 
$\vec\nabla\times \vec v$ becomes singular (behaves as a Dirac delta function,
see for example \cite{andron}).
We refer to $n$ as a topological charge since any continuous deformation of 
the contour $C$, which does not incorporate any other points where
curl of the velocity field does not vanish, can not change $\Gamma_C$. 

\section{The Helmholtz-Kelvin theorem}
\label{dowod}
One may ask whether the circulation $\Gamma_C$ is conserved in time evolution 
of the
velocity field. For a fixed contour, a possibly moving vortex may leave an area 
bounded by the contour and consequently $\Gamma_C$ changes its value. 
Thus, the relevant situation we should consider
corresponds to the case 
where we let an initially defined contour evolve in time 
according to the velocity field. 

Parameterizing the initial contour by $\xi$ variable, i.e. 
$C(t_0)=\{\vec{r}(\xi,t_0)\}$, and employing the equation of motion of a contour
\begin{equation}
\frac{d}{dt}\vec{r}(\xi,t)= \vec{v}[\vec{r}(\xi,t)],
\label{motion}
\end{equation}
yields
\begin{equation}
\frac{d}{dt}\Gamma_{C(t)} = \oint_{C(t)} \left(
\frac{\partial\vec{v}}{\partial t} + (\vec{v}\cdot \vec{\nabla})\vec{v}
+ \frac{1}{2}\vec{\nabla}v^2 \right)\cdot d\vec{l}.
\label{dcirc}
\end{equation}
If the contour is drawn through points where the velocity field $\vec v$ and its
partial derivatives are well defined, applying: Eq.~(\ref{dyna}), equality 
$(\vec{v}\cdot \vec{\nabla})\vec{v}=\frac{1}{2}\vec{\nabla}v^2 - \vec{v}
\times (\vec{\nabla} \times \vec{v})$,
and using the
fact that, at least on the contour $\vec{\nabla} \times \vec{v} = 0$, we get  
\begin{equation}
\label{gamma0}
\frac{d}{dt}\Gamma_{C(t)} = 0.
\end{equation}
Equation (\ref{gamma0}) establishes the Helmholtz-Kelvin theorem 
originally proved for a non-viscous fluid \cite{landau}. Applicability of this
theorem to a non-linear quantum systems 
is presented in the Appendix.

The HKT is fulfilled provided the contour $C(t)$ that goes initially through 
points where the velocity field is well defined will evolve in time through 
such points only. In the following we will see how such an assumption may be 
broken by quantum evolution of the wave function
leading to a change of the topological charge. 

\section{Analytical examples}
\label{przyklady}
In this section we discuss a few possibilities of breaking of the HKT. 
To this end we analyze three different analytically solvable examples where
breaking of the HKT becomes evident. 

First example shows that changes of the topological charge of a vortex 
correspond to the appearance of a nodal line of the wave function. 
In Ref.~\cite{ripoll} the authors show the instability of a vortex 
placed in an anisotropic two-dimensional harmonic trap.
We would like to discuss 
the source of violation of the HKT in such a case.
We will calculate the velocity field 
and analyze 
changes of its circulation from a point of view of the HKT.

Consider an anisotropic two-dimensional harmonic oscillator with the
potential 
\begin{equation}
V(x,y)=\frac{1}{2}x^2 + \frac{\lambda^2}{2} y^2,
\label{hpot}
\end{equation}
(where we use the units of the harmonic oscillator corresponding to
the $x$-direction). 
The initial wave function 
is prepared as a superposition of the two lowest excited states
\begin{equation}
\Psi(x, y) \propto (x + i \alpha y ) 
e^{-(x^2 + \lambda y^2)/2} ,
\label{initial1}
\end{equation}
with a real parameter $\alpha>0$.
Choosing any contour $C$ that encircles the origin of the coordinate frame 
we find out that the circulation 
$\Gamma_C$ corresponds to $n=1$ vortex.
However, the circulation 
around that point is not conserved in time. Indeed, analysis of 
time evolution of the velocity field
\begin{equation}
\vec{v}(x,y,t)=
\frac{\alpha \ \cos(E t)}{
x^2 + y^2 \  \alpha^2 + 2 \alpha x y \sin(E t)}
(-y \vec{e}_x+ x\vec{e}_y),
\label{vel}
\end{equation}
(where $E=\lambda-1$ 
is the energy difference of the two lowest excited states)
reveals that for $Et \in [0, \pi/2)$,  
$\Gamma_C$ corresponds to $n=1$
while
for $Et \in (\pi/2, \pi]$ to $n=-1$.

To discuss applicability of the HKT 
we have to 
investigate time evolution of the initially defined contour.
The problem requires careful treatment because
for $Et=\pi/2-\varepsilon$ (where $\varepsilon\rightarrow
0^+$) the velocity field goes to zero everywhere
except on the nodal line, $x+\alpha y=0$, 
where it diverges
$\vec{v} \rightarrow \frac{1}{\varepsilon x}(\vec e_x +\alpha \vec e_y)$.
To demonstrate violation of the HKT we have 
to show that a contour initially encircling the center 
is not extended to infinity but indeed faces singularity for $Et=\pi/2$.

\begin{figure}
\centering
{\includegraphics[clip = true, width=7.6cm]{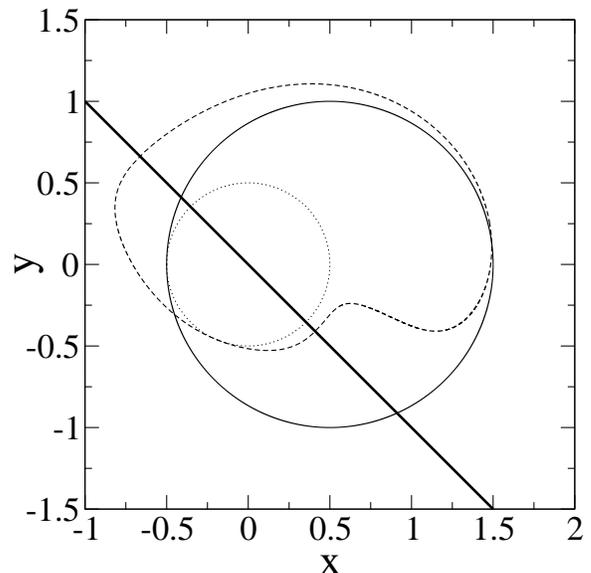}}
\caption{Plot of the contour encircling the center of
the trap for $Et \rightarrow \pi/2$ (dashed line). The contour was initially
($t=0$) chosen to be a circle with the unit radius centered at $(0.5, 0)$
(solid line).
The thick solid line represents the nodal line given by the equation
$x+\alpha y=0$.
Initially, the point of the contour closest to the origin of the coordinates
is situated at the distance $0.5$. Therefore in course of the evolution 
none of the points of the contour 
can get closer to the origin than at the distance 0.5, i.e.
can not cross the dotted circle with the radius
$0.5$ that is centered at $(0,0)$. For this presentation we have chosen
$\lambda=\sqrt{2}$, $\alpha= 1$ --- see text for details.
}
\label{contour}
\end{figure}

Suppose we define a contour, encircling the origin,
at a time $t=0$, and look for a time evolution 
of its arbitrary point $i$, whose position we denote by 
$\{x_i(t), y_i(t)\}$. Combining Eq.~(\ref{motion}) and Eq.~(\ref{vel}) we get
\begin{eqnarray}
\frac{dx_i}{dt} = -f(x_i,y_i)\;y_i, \ \ \ \frac{dy_i}{dt} =f(x_i,y_i)\; x_i,
\label{ewolcontour}
\end{eqnarray}
with
\begin{equation}
f(x,y)=\frac{\alpha \cos(E t)}{x^2 + y^2 \  \alpha^2 + 2 \alpha x y \sin(E t)}.
\label{funkcja}
\end{equation}
From  Eq.~(\ref{ewolcontour})
it is apparent that the distance of the point $\{x_i(t), y_i(t)\}$
from the origin does not change in course of the evolution. 
Indeed the quantity 
$x_i(t)^2+ y_i(t)^2$ is a constant of motion
as long as the function (\ref{funkcja})
is well defined. Note that it diverges,  at $Et=\pi/2$,  for points
lying on the nodal line. Therefore, for
$Et=\pi/2-\varepsilon$ with arbitrary small  $\varepsilon>0$
nothing extraordinary happens with the contour until at $Et=\pi/2$ the
contour faces singularity (see Fig.~\ref{contour} for an example of the 
contour evolution).
Thus, the HKT can not be applied because 
quantum evolution 
of the wave function pushes an evolving contour to singular points and 
the quantities involved in the theorem become
undefined.

We would like to stress that the presented scenario where the vortex situated in
the harmonic trap reveals periodic changes of the topological charge is possible
only when the initial vortex is placed exactly at the center of the potential.
Otherwise the vortex changes its position --- it escapes to infinity and another
vortex with the opposite charge arrives from infinity and the scenario repeats
periodically.

In the second example we would like to present violation of the HKT 
due to annihilation of a vortex ring.
This example
comes from Ref.~\cite{bial00} where creation and annihilation
of a vortex ring for freely moving particle is
presented.
Even though it is not clearly stated in Ref.~\cite{bial00},
the HKT is not fulfilled in that case.
The wave function of a vortex ring \cite{bial00}
(in dimensionless units)  may be written in the form
\begin{eqnarray}
\Psi(x, y, z, t) & \propto & 
[(x-kt)^2 + y^2 + z^2 - 1 \nonumber\\  
& + & i 3 (z + t)]
e^{ikx-i k^2 t/2},
\end{eqnarray}
where the wave vector $\vec{k}=(k,0,0)$ is related to a motion of the ``center 
of mass'' of the vortex ring with a constant velocity $\vec{k}$.
The vortex ring corresponds to the nodal line of the wave function and 
is located at the intersection of the plane $z+t=0$ and sphere 
$(x-k t)^2 + y^2 + z^2 = 1$. At time $t=-1$, the vortex  is born 
at a point $(-k,0,1)$ and, at time $t=1$, it disappears at a point $(k,0,-1)$. 
The radius of the vortex changes in time as $\sqrt{1 - t^2}$.

\begin{figure}
\centering
{\includegraphics[clip = true, width=4cm]{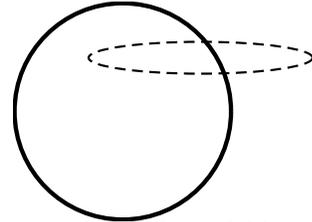}}
\caption{ Schematic plot of the
nodal line corresponding to a vortex ring (solid line) 
and a contour that encircles the ring (dashed line). 
In last stage of time evolution the ring
shrinks to a point and the contour faces singularity.}
\label{fig2}
\end{figure}

Suppose at any time $t \in (-1, 1)$, we define a contour $C$ so that 
it encircles the vortex (see Fig.~\ref{fig2})
and the circulation of the velocity field
corresponds to $n=1$. The evolving contour can not cross the vortex ring 
without facing a singularity in the velocity field. However, the ring 
at some moment starts shrinking and at $t=1$ it reduces to a point.
Consequently, at $t=1$, the contour must go through a singularity of 
the velocity field and the integral (\ref{dcirc}) and also the HKT 
become meaningless.

The last example we would like to comment on shows collision of
two vortices leading to the appearance of a single doubly charged vortex.
Consider, for simplicity, a two-dimensional H atom 
initially in the first excited state with angular momentum $L_z=1$
[i.e. $\psi_1(r,\varphi)\propto r\exp(-2r/3)\exp(i\varphi)$ in atomic units]
that is driven resonantly by a circularly polarized electromagnetic field. 
The field frequency  is tuned to the transition between the first and
second excited energy eigenstates. 
It results in the familiar Rabi oscillation \cite{cohen} 
between the $\psi_1(r,\varphi)$ state
and the second excited state with angular momentum $L_z=2$
[i.e. $\psi_2(r,\varphi)\propto
r^2\exp(-2r/5)\exp(i2\varphi)$].
Indeed, the time evolution of the wavefunction reads
\begin{equation}
|\psi(t)\rangle \propto 
\cos(Dt) e^{-i E_1 t}\;|\psi_1\rangle+e^{-i E_2 t}\sin(Dt)\;|\psi_2\rangle\
,\end{equation}
where $D$ is a dipole matrix element \cite{cohen}; $E_1$, $E_2$ are energies 
of $|\psi_1\rangle$, $|\psi_2\rangle$ states.

For $t=0$ there is a vortex with $n=1$ at the center of the coordinates 
as expected for the state with angular
momentum $L_z=1$.
When $Dt$ increases another vortex with $n=1$ moves in from {\it infinity}, 
collides with the first one (situated at the center of the coordinates during
the whole time evolution)
at $Dt=\pi/2$  and then moves out to 
{\it infinity} again and so on. 
During the collision, i.e. for $Dt=\pi/2$, a single vortex with $n=2$
is formed. 
In the considered example, the HKT does not apply: if we define a contour so
that it encircles the vortex with $n=1$ situated at the center of the 
coordinate, such a contour encounters a singularity in the 
velocity field during  the collision with the other vortex. Such a behavior,
is expected to hold in an arbitrary collision between two vortices leading to
formation of one, doubly charged, vortex.

These examples are not the only ones   illustrating 
that vortices can disappear
or change their charge in the course of time evolution.
They were 
chosen to illustrate that such  processes can happen 
in different physical systems and results in the violation of the HKT.
The theorem is explored in the context of vortices' stability
(see e.g. ~\cite{anderson})
to imply the constancy of a vortex topological charge. As we have 
clearly 
illustrated the HKT can not assure persistence of vortex currents in the
quantum mechanical systems. Therefore we conclude that 
the vortex topological charge is not as robust quantity as it is commonly 
believed (see e. g.  \cite{ripoll} and references therein, \cite{coullet}).
Finally, we would like to mention that appearance of vortices can happen 
in a way that does not violate the HKT. Namely, they can appear in the form 
of a closed vortex line that springs from a point or as a vortex-antivortex 
pair creation from a node of a wave function \cite{bial00}. 
\section{Summary and conclusions}
We have analyzed the applicability of the Helmholtz-Kelvin theorem in the
hydrodynamical formulation of the quantum mechanics. 
The velocity field of the probability fluid is defined as the
gradient of the phase of the quantum
wave function. This implies that nonzero circulation, 
along a given contour, may come out in the system
if the field reveals a singularity at certain points on a 
surface encircled by the contour. 
Adopting the HKT to the quantum liquid 
may suggest that such a topological charge of the system can not change. 

However, the HKT may be employed if a given contour evolves through points
where the velocity field is well defined. It may happen even in classical
hydrodynamics that such an assumption is not fulfilled if, e.g., liquid
encounters obstacles in the flow \cite{landau}. 
In quantum liquids the situation is more complicated because a singularity is
necessary for a nonzero circulation. Indeed, phase of a wave function is
undefined at a vortex core, which means that singularities appear in 
hydrodynamical formulation of quantum mechanics whenever vortices show up. 
This property 
makes distinction of hydrodynamical description of quantum system from 
the classical hydrodynamics.
We have presented simple analytical examples where the quantum 
evolution of a wave function pushes a contour to a singular point. 
Such a process is accompanied by a change of the vortex topological charge.

We have illustrated the violation of the HKT choosing examples that 
correspond to linear quantum systems. Macroscopic quantum behavior is present 
in interacting many particle systems that are usually described in the mean 
field approximation by a nonlinear Schr\"odinger equation. In Appendix we
complete the analysis of the HKT for a nonlinear Schr\"odinger evolution. 
Vortices maybe also investigated in propagation of {\it classical} 
light \cite{coullet,bambrilla}. Indeed, the evolution equation of the 
slowly varying envelope of a 
light beam can have identical form as the Schr\"odinger
equation, which allows for exactly the same considerations as we have 
presented above. The dynamics of vortices 
and escapes of the off-center vortex in an analog of the asymmetric 
harmonic potential
have been experimentally observed in such systems \cite{light}.

We are grateful to J. Zakrzewski and G. Molina-Terriza for fruitful discussion 
and to U. Fischer for
the critical reading of the manuscript.
Support of KBN under project 5~P03B~088~21 is acknowledged.

\section{Appendix: Helmholtz-Kelvin theorem for nonlinear systems}
\label{nonlinear}

For completeness we analyze here the application of 
the HKT in hydrodynamical formulation 
of nonlinear quantum mechanics. 
The so-called Gross-Pitaevskii equation is successfully used to describe 
the properties of a Bose-Einstein
condensate in trapped alkali atoms \cite{pitaevskii}. 
In particular this equation exhibits various vortex solutions \cite{anderson}. 
To make the discussion more general
we will also include the mean field term describing a possible dipolar 
interactions between condensed atoms \cite{goral}.
The nonlinear Schr\"odinger equation of interest reads
\begin{eqnarray}
\label{ham1}
i\hbar\frac{\partial}{\partial t}\Psi(\vec{r},t)
= &-& \frac{\hbar^2}{2m} \vec{\nabla}^2\Psi(\vec{r},t)+ V(\vec{r},
t)\Psi(\vec{r},t)\nonumber\\
&+& \left( \int d^3r' V_{int}(\vec{r}-\vec{r}') |\Psi(\vec{r}',t)|^2 \right) \Psi(\vec{r},t),
\end{eqnarray}
where $V_{int}$ corresponds to two body interactions (e.g. point or dipolar
interactions \cite{pitaevskii,goral}). Substitution 
$\Psi(\vec{r}, t)=\sqrt{\rho(\vec{r}, t)} \exp(i \chi(\vec{r}, t))$ 
leads to the continuity equation (\ref{cont}) and the following 
dynamical equation for the velocity field 
\begin{eqnarray}
\label{dyna1}
 m \frac{\partial \vec{v}}{\partial t} &+&
 \vec{\nabla} \left( \frac{1}{2} m v^2 + V+ \int d^3r' V_{int}(\vec{r}-\vec{r}') \rho(\vec{r}',t)\right.  \nonumber\\  
 &-& \left. \frac{\hbar^2}{2 m}\frac{\nabla^2 \sqrt{\rho}}{\sqrt{\rho}}
 \right)  = 0.
\end{eqnarray}
The only difference in comparison with Eq.~(\ref{dyna}) is the appearance
of two new terms in the bracket of Eq.~(\ref{dyna1}). These terms, however,
do not change the proof of the HKT --- 
for example Eq.~(\ref{dcirc}) does not possess any potential like terms. 
Therefore the Helmholtz-Kelvin
theorem holds also for systems described by Eq.~(\ref{ham1}) under
the same assumptions as in the case of linear quantum mechanics
discussed in Sec.~\ref{dowod}.


\begin{references}

\bibitem{feyn} R. P. Feynman, R. B. Leighton, M. Sands, 
The Feynman Lectures on Physics, Volume II,
(Addison-Wesley Publishing Company, Massachusetts, 1963).  

\bibitem{ghosh} S. K. Ghosh, B. M. Deb, Phys. Rep. {\bf 92} 1 (1982).

\bibitem{bial92} I. Bia\l{}ynicki-Birula, M. Cieplak, and J. Kaminski, Theory of Quanta
(Oxford University Press, Oxford, 1992).

\bibitem{bial00} I. Bia\l{}ynicki-Birula, Z. Bia\l{}ynicka-Birula, and C. \'Sliwa,
Phys. Rev. A {\bf 61} 032110 (2000).

\bibitem{andron} E. L. Andronikashvili and Yu. G. Mamaladze, Rev. Mod. Phys. {\bf 38} 567 (1966).

\bibitem{cornell} M. R. Matthews {\it et al.}
Phys. Rev. Lett. {\bf 83}, 2498 (1999).

\bibitem{dalibard} K. W. Madison, F. Chevy, W. Wohlleben, and J. Dalibard, Phys. Rev. Lett. {\bf 84}, 806 (2000).

\bibitem{pitaevskii}  F. Dalfovo, S. Giorgini, L. P. Pitaevskii, and 
S. Stringari Rev. Mod. Phys. {\bf 71} 463 (1999).  

\bibitem{landau} L. D. Landau, E. M. Lifshitz, Fluid Mechanics
(Butterworth-Heinemann, 1995). 

\bibitem{bolda} E. L. Bolda and D. F. Walls, Phys. Rev. Lett. {\bf 81} 5477
(1998).

\bibitem{anderson} B.P. Anderson {\it et al.}
Phys. Rev. Lett. {\bf 86} 2926 (2001).

\bibitem{koplik} J. Koplik and H. Levine, Phys. Rev. Lett. {\bf 71} 1375
(1993).

\bibitem{feyn1} R. P. Feynman, Statistical Mechanics, (W. A. Benjamin, Massachusetts, 1972). 

\bibitem{ripoll} J. J. Garc\'{\i}a-Ripoll, G. Molina-Terriza, V. M. P\'{e}rez-Garc\'{\i}a,
and L. Torner, Phys. Rev. Lett. {\bf 87} 140403 (2001).


\bibitem{cohen} C. Cohen-Tannoudji, J Dupont-Roc, G. Grynberg, Atom-Photon
Interactions: Basic Processes and Applications (John Wiley $\&$ Sons, 1992).

\bibitem{coullet} P. Coullet, L. Gil and F. Rocca, Opt. Comm. {\bf 73} 403
(1989).

\bibitem{bambrilla} M. Brambrilla {\it et al.} Phys. Rev. A {\bf 43} 5090
(1991).

\bibitem{light} G. Molina-Terriza, E. M. Wright, and L. Torner, Opt. Lett. {\bf
26} 163 (2001); G. Molina-Terriza, J. Recolons, J. P. Torres, and L. Torner, 
Phys. Rev. Lett. {\bf 87} 023902 (2001). 


\bibitem{goral} K. G\'oral, K. Rz\c a\.zewski and T. Pfau, Phys. Rev. A 
{\bf 61} 051601(R) (2000).

\end{references}
\end{document}